\documentclass[prb,showpacs,preprintnumbers,amssymb,eqsecnum,twocolumn]{revtex4}
\usepackage{graphicx}
\usepackage{dcolumn}
\usepackage{bm}
\def\be{\begin{equation}}
\def\ee{\end{equation}}
\def\bea{\begin{eqnarray}}
\def\eea{\end{eqnarray}}

\begin{document}

\title{Conductance distribution in three dimensions: analytic solution of the Generalized DMPK equation in the strongly disordered regime}
\author{Andrew Douglas and K. A. Muttalib}
\affiliation{Department of Physics, University of Florida, P.O. Box 118440,
Gainesville, FL 32611-8440}
\begin{abstract}
We develop a systematic perturbative method to obtain analytic solution of the Generalized Dorokhov-Mello-Pereyra-Kumar (DMPK) equation in the strongly disordered regime which describes the evolution of the joint probability distribution of the transmission eigenvalues with system size. The solution allows us to obtain the distribution of conductance analytically in the insulating regime. Our results are consistent with existing numerical simulations of the three dimensional tight binding Anderson model, and suggests a possible description of the Anderson transition in the presence of a very broad, highly asymmetric conductance distribution. 

\end{abstract}


\maketitle

\section{Introduction}

The apparent existence of metal-insulator transitions in 
two-dimensions \cite{2d} has generated a lot of interest recently on the 
problem of interplay between interaction and disorder. However, the consequences of a highly non-trivial, non-Gaussian distribution of conductances $P(g)$ for even non-interacting 
electrons at arbitrary disorder at zero temperature in two or three dimensions remains poorly understood. In particular, several novel features in $P(g)$ have recently been discovered 
\cite{mwgg,mu-wo,gmw,mwg,martin,plerou,perez,poirier,mohanty,mmwk,mmw}, 
including a non-analyticity in $P(g)$ near $g=1$ in the insulating regime 
\cite{mwgg}, a `one-sided' log-normal distribution \cite{mu-wo} near the
metal-insulator crossover regime in quasi one dimension (Q1D) as
well as strong deviations from the expected log-normal
distribution for an Anderson insulator \cite{mmwk,mmw} in three
dimensions (3D). They raise serious questions about the effect of such broad asymmetric distributions on e.g. the Anderson metal-insulator transition in particular and on quantum phase transitions \cite{vojta} in general.

Although there is a significant body of numerical work available \cite{markos,soukoulis,slevin},
analytic study of the full distribution $P(g)$ in 3D at arbitrary disorder is
beyond the scope of the conventional field theory framework \cite%
{altshuler,shapiro}. The novel features in $P(g)$ mentioned above were obtained directly  
from the joint probability distribution $p(\textbf{x})$ of the $N$ transmission 
eigenvalues $\textbf{x}\equiv \{x_a\}$, $x_a >0$, for an $N$-channel disordered conductor, via the 
Landauer formula\cite{landauer}
  \be \label{landauer}
  P(g)\propto \int\prod_a^N dx_a p(\textbf{x})\delta
  \left(g-\sum_i \rm{sech}^{2}x_i\right).
  \label{Pofg-def}
  \ee 
The N-dimensional integral was evaluated using a saddle point method \cite{mu-wo,gmw}, while the distribution $p(\textbf{x})$ was obtained from the Q1D DMPK equation \cite{dmpk} and its 3D generalization obtained recently \cite{mk,mg}. For a conductor of fixed cross section $L^2$, it describes the evolution with
length $L_z$ of the joint probability distribution $p_{L_z}(\textbf{x})$
based on a `local' maximum entropy ansatz \cite{smmp}. The generalized DMPK (GDMPK) 
equation reads \cite{mk,mg}
  \begin{eqnarray}\label{gdmpk}
  \frac{\partial p(\textbf{x},t)}{\partial t} = \frac{1}{4} 
  \sum_{i=1}^N \frac{\partial}{\partial x_i}K_{ii}\left[ \frac{\partial}{\partial x_i}+
  \frac{\partial \Omega}{\partial x_i}\right]p(\textbf{x},t) 
  \end{eqnarray}

where
  \bea
  \Omega  \equiv  &-&\sum_{i<j} \gamma_{ij} \ln |\sinh^2 x_j-\sinh^2 x_i|\cr 
  &-&\sum_i \ln |\sinh 2x_i |.
  \eea
In the above, $t\equiv L_z/l$ where $l$ is the mean free path, $\gamma_{ij}\equiv 2K_{ij}/K_{ii}$ where
$K_{ij}$ is a phenomenological matrix \cite{mg} defined in terms of certain eigenvector 
correlations that can be explicitly evaluated numerically and contains information about dimensionality. 
The Q1D DMPK equation is recovered when $\gamma_{ij}\to 1$ (we only consider orthogonal symmetry). 

Recently a direct numerical solution of Eq.~(\ref{gdmpk}) has been obtained
\cite{markos1} by mapping it onto a Langevin equation. The numerical results confirm that the GDMPK equation indeed distinguishes between two and three dimensional models with the same number of transmission channels. With detailed comparisons, the study confirms that: (i) Eq.~(\ref{gdmpk})
correctly describes the full $P(g)$ not only in the strongly localized
regime but also near the critical regime in 3D, and (ii) two parameters, $%
K_{11}$ and $K_{12}$, are enough to model the entire matrix $K_{ab}$ at least in these
regimes. In Ref.~[\onlinecite{mk}] it was shown that at least two parameters are needed in order to guarantee the conservation of probability, so this can be considered as a minimalistic model. Within this approximation, disorder is characterized by $\Gamma \equiv l/K_{11}L_z$.
In the insulating regime one can interpret $\xi \equiv 4l/K_{11}$ as the localization 
length, giving $\Gamma = \xi/4L_z$. The other independent parameter
$\gamma_{12} = \xi/8L$, so that $\Gamma/\gamma_{12}=2L/L_z$ depends only on the geometry of the system. Therefore for a cubic system, within this approximation, the entire distribution is characterized by a single disorder parameter $\Gamma$.

As shown in Refs.~[\onlinecite{mmwk,mmw}]  the parameter $K_{11}(L)$, where $L$ is the cross-sectional dimension, contains information about the dimensionality of the system via the dimension-dependent eigenvector correlations that define the matrix $K$. In particular in 3D,

  \bea
	K_{11} \propto && 1/L^0, \;\;\;\; \rm{insulating \;\;regime,} \cr
	&& 1/L^1,  \;\;\;\; \rm{at \;\;critical \;\;point,}\cr
	&& 1/L^2, \;\;\;\; \rm{metallic\;\; regime,}
	\eea
such that the quantity $\tilde{K}_{11}\equiv \lim_{L\to \infty}K_{11}(L)$ is zero in the metallic regime as well as at the critical point, but it is finite for insulators. Clearly the parameter  $\tilde{K}_{11}$ can be considered as an order parameter for the Anderson transition. We can therefore expect that Eq.~(\ref{gdmpk}) can be used as the
starting point for studying the Anderson transition in terms of the full
distribution of conductances, starting from the insulating limit. However, unlike the Q1D DMPK equation, 
a general analytic solution of the GDMPK equation for arbitrary disorder 
is not yet available. 

In this paper we will obtain an analytic perturbative solution of the GDMPK equation for $N$ transmission channels valid in the strongly disordered regime $\Gamma \ll 1$. The solution is based on a mapping of the equation on to an imaginary time Schr\"odinger equation \cite{beenakker}, describing $N$ interacting bosons in one dimension evolving from a delta function initial condition and satisfying some singular boundary conditions. In Ref.~[\onlinecite{mmw}], the full $P(g)$ in 3D was evaluated based on a solution of the GDMPK equation in the limit where the interaction between the eigenvalues was totally neglected. In this work, we include the interaction and show that it changes the qualitative features of the distribution. We begin by developing a systematic perturbation theory method to obtain the $N$-particle Greens function. As a first check, we show that our solutions agree where exact solutions are available, namely for the special cases of $\gamma_{12}=1,2,4$ \cite{beenakker}. We then use the $N$-particle Greens function to obtain the joint probability distribution of the eigenvalues $x_a>0$, $a=1,2\cdots N$, analytically. 
We show that in particular, the density of the eigenvalues remains uniform, independent of disorder, with an exponential gap at the origin that increases with increasing disorder.  Indeed, it is this feature which is at the heart of many novel properties in the distribution of conductances in 3D. It suggests that the Anderson metal to insulator transition can be viewed as the opening of a finite gap in the density of the eigenvalues at the origin. Finally, we do the $N$-dimensional integral of  Eq.~(\ref{Pofg-def}) approximately to show that the distribution of conductance agrees with the numerical results, improving on the approximate solution obtained in Ref.~[\onlinecite{mmw}] where the distribution was found to be much broader. In particular, we obtain the variance $ \sigma(\ln g)$ as well as the skewness $ \chi(\ln g)$. We find that asymptotically the variance is not a simple power law although numerical data can be fitted with a power law in limited regimes.  We also find that asymptotically the skewness goes as a positive constant.  Further, the skewness changes sign from positive to negative well before the metal-insulator transition point.  A short version containing some of the above results was published in Ref.~[\onlinecite{dm-09}]. Here we provide details of the calculation and also these new results. 

The paper is organized as follows. After a brief introduction in section I, we discuss the mapping of the GDMPK equation on to a Schr\"odinger Equation in section II. In section III we show how a systematic perturbation theory can be developed and point out how a resummation of the perturbative expansion is essential in order to satisfy certain boundary conditions. We will use the results obtained in section III to obtain the joint probability distribution of transmission eigenvalues in section IV and then the full distribution of conductances in section V. Section VI contains summary and conclusions. In Appendix A we discuss how to include systematic corrections to the leading order results of section III.

\section{Mapping GDMPK on to a Schr\"odinger Equation}

Within the approximation that only two independent matrix elements $K_{11}$ and $K_{12}$ can describe the entire $K$-matrix, Eq.~(\ref{gdmpk}) can be written in the form
	\be
	\frac{\partial p(\textbf{x},t)}{\partial t}= \frac{K}{4} 
	\sum_{i=1}^N \frac{\partial}{\partial x_i}\left[ \frac{\partial}{\partial x_i}+
	\frac{\partial \Omega}{\partial x_i}\right]p(\textbf{x},t)
	\ee

	\bea
	\Omega\equiv &-& \gamma\sum_{i<j} \ln | \sinh^2 x_j-\sinh^2 x_i |\cr
	&-& \sum_i \ln |\sinh 2x_i |.
	\eea
where $K \equiv K_{11}$, and $\gamma \equiv \gamma_{12}$.  The initial condition is given by:
  \be
   p(\textbf{x},t=0)=\delta(\textbf{x})
  \ee 
and the boundary conditions are
  \be
  \lim_{x_i\rightarrow 0} \left[\frac{\partial}{\partial x_i}+
  \frac{\partial \Omega}{\partial x_i}\right]p =0;\;\;\; \lim_{x_i\rightarrow \infty}p=0.
  \ee
Following Ref.~[\onlinecite{beenakker}] we map the equation for $p$ on to an imaginary time
Schr\"odinger equation. First we use a factorization
  \be\label{pofxyt}
  p(\textbf{x},t|\textbf{y})=\xi(\textbf{x})G_N(\textbf{x};t|\textbf{y})\xi^{-1}(\textbf{y}); \;\;\; \xi\equiv   
  e^{-\Omega/2}.
  \ee
The $N$-particle Greens function $G_N(\textbf{x};t|\textbf{y})=G(x_1,x_2, \cdots ,x_N;t|y_1,y_2, \cdots ,y_N)$ then satisfies the imaginary time Schr\"odinger equation in the variables $x_i$, given by
  \be
  -\partial G_N/\partial t=HG_N
  \label{eqofGN}
  \ee 
where
	\begin{equation}  \label{ham}
	H=\sum_i \left[-\frac{K}{4}\frac{\partial^2}{\partial x^2_i} 
	+ u(x_i)\right]+\sum_{i<j}v(x_i,x_j).
	\ee
and the single particle potential and the interaction terms are given by
	\begin{eqnarray}
	\label{potentials}
	u(x_i) &\equiv& \frac{\lambda_u}{\sinh^2 2x_i}\cr
	v(x_i,x_j) &\equiv& \lambda_v [
	\frac{1}{%
	\sinh^2(x_i-x_j)}\cr
	&+&\frac{1}{\sinh^2(x_i+x_j)}].
	\end{eqnarray}
Here 
	\bea
	\lambda_u &\equiv & -K/2 \cr
	\lambda_v &\equiv & K\gamma(\gamma-2)/8
	\eea 

The Greens function $G_N(\textbf{x};t|\textbf{y})$ will then have the initial condition 
	\be
	G_N(\textbf{x}; t=0|\textbf{y}) = \frac{1}{N!}\sum_{\pi(y)}\epsilon_{S,AS}(\textbf{y}) \prod_{i=1}^N \delta(x_i-y_i)
	\ee
where the subscripts $S$ and $AS$ refer to the symmetric and antisymmetric initial conditions, respectively, depending on whether $\Omega$ is symmetric or antisymmetric in its variables and $\pi$ stands for permutations. In the small $y$ limit, this recovers the proper initial conditions on $p$. The boundary conditions become
	\be
	\lim_{x_i\rightarrow 0}\left[\frac{\partial G_N}{\partial x_i}-\frac{G_N}{\sinh 2x_i}\right] = 0.
	\ee
Note that the factorization requires defining a new set of variables $y_i$, but in the end we will be interested in the $y\to 0$ limit 
	\be
	p(\textbf{x},t)=p(\textbf{x},t|\textbf{y}=0).
	\ee
Therefore, we will be interested in the $y\to 0$ limit of the $N$-particle Greens function $G_N(\textbf{x};t|\textbf{y})$ as well. According to the construction of Eq.~(\ref{pofxyt}), 
$p(\textbf{x},t|\textbf{y})$ looks highly singular in the limit $y\to 0$, given by
	\begin{eqnarray}\label{singular}
	p(\textbf{x},t|\textbf{y}) &=&
	\frac{\prod_{i<j}|\sinh^2x_j-\sinh^2x_i|^{\frac{\gamma}{2}} 
	\prod_i|\sinh 2x_i|^{\frac{1}{2}}}
	{\prod_{i<j}|\sinh^2y_j-\sinh^2y_i|^{\frac{\gamma}{2}}
	\prod_i|\sinh2y_i|^{\frac{1}{2}}}\cr
	& \times & G_N(\textbf{x},t|\textbf{y}).
	\end{eqnarray}
However, it turns out that the complete $N$-particle Greens function can be obtained exactly for the special values $\gamma = 1, 2, 4$. In these special cases \cite{beenakker} the small $y_m$ limit of $G_N$ has the form 
	\begin{eqnarray}\label{symmetry}
	G_N(\textbf{x},t|\textbf{y}) &\propto & \prod_{i<j}|\sinh^2y_j-\sinh^2y_i|^{\frac{\gamma}{2}}\cr
	&\times & \prod_i|\sinh 2y_i|^{\frac{1}{2}}; \;\;\; y_m \ll 1.
	\end{eqnarray}
Thus the $G_N$ for these $\gamma$ cancel the singularity arising from $\xi^{-1}(\textbf{y})$ exactly as $y_m\rightarrow 0$ and yields a finite result for $p(\textbf{x},t)$. This is a highly non-trivial result, and typically an approximate evaluation of the $N$-particle Greens function will not have the same property, rendering the extraction of $p(\textbf{x},t)$ impossible.  In what follows, we will organize the perturbative series for $G_N$ in such a fashion as to conform to this symmetry.  

\section{The Greens function}

As given in Eq.~(\ref{eqofGN}), the $N$-particle Greens function satisfies a differential equation which is different from the usual many-body Greens function defined in terms of time-ordered annihilation and creation operators. In fact it satisfies the differential equation obeyed by N-particle propagator. Taking a cue from this, one can straightforwardly verify that $G_N$ can be written as:
	\begin{eqnarray}
	G_N(\textbf{x},t|\textbf{y}) = \frac{1}{N!}\langle \textbf{0}|\prod_{m=1}^N\psi(x_m) e^{-Ht}
	\prod_{n=1}^N \psi^{\dagger}(y_n)|\textbf{0}\rangle
	\label{Prop}
	\end{eqnarray}
where $|\textbf{0}\rangle$ is the $N$-particle vacuum and 
	\begin{eqnarray}
	H &=& \int_{-\infty}^{\infty}dx \psi^{\dagger}(x)
	\left[-\frac{K}{4}\frac{\partial^2}{\partial x^2}+u(x)\right]\psi(x) \cr
	&+& \frac{1}{2}\int_{-\infty}^{\infty}dx\int_{-\infty}^{\infty}dx' \psi^{\dagger}(x) \psi^{\dagger}(x')\cr
	&\times& v(x,x')\psi(x') \psi(x). 
	\end{eqnarray}
This marks a distinction between the $G_N$ of Eq.~(\ref{Prop}) and more typical many-body Green's functions in that the expectation in this case is against the vacuum, rather than the many-body ground state.  As we will see, this will have dramatic consequences for the diagrammatic expansion of $G_N$.  Now as usual we would like to go to the interaction picture, and so we define the interaction picture imaginary time evolution operator:
	\begin{eqnarray}
	S(t_2 ,t_1 ) = T\exp \left\{ { - \int\limits_{t_1 }^{t_2 } {dt'\,H'} (t')}  \right\}
	\end{eqnarray}
where $H'$ is the perturbation in the interaction picture. Taking advantage of the fact that $H'$ annihilates the vacuum, and that t is always greater than 0, we can write Eq.~(\ref{Prop}) as:
	\bea
	G_N ({\bf{x}},t|{\bf{y}}) &=& \frac{1}{{N!}}\left\langle {\bf{0}} \right|TS(\infty ,0)\prod\limits_{m = 1}^N {\psi (x_m 	,t)} \cr
	&\times&\prod\limits_{n = 1}^N {\psi ^\dag  (y_n ,0)} \,\left| {\bf{0}} \right\rangle 
	\eea
From the standard theory of non-equilibrium Green's functions, the diagrammatic expansion of $G_N$ follows. We would compute all diagrams with N external legs $x_i$, and N external legs $y_i$ as illustrated in Fig.~\ref{bubble}.  
	\begin{figure}
	\includegraphics[angle=0,width=0.35\textwidth]{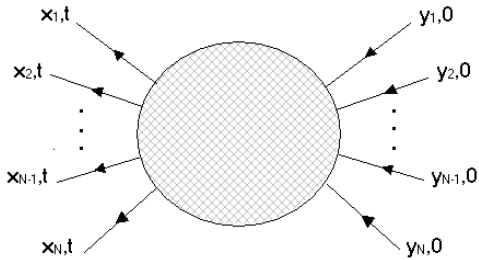}
	\caption{Diagrammatic expansion of $G_N$}
	\label{bubble}
	\end{figure}
And we would connect them via the external potential and interaction vertices shown in Fig.~\ref{vertices}. 
	\begin{figure}
	\includegraphics[angle=0,width=0.45\textwidth]{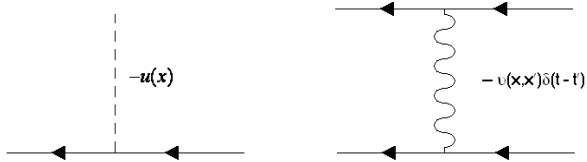}
	\caption{Vertices appearing in diagrammatic expansion of $G_N$}
	\label{vertices}
	\end{figure}
Finally, the unperturbed Green's function would be given by:
	\begin{eqnarray}
	G_0 (t_1 ,t_2 ) =  - \left\langle 0 \right|T\psi (t_1 )\psi ^\dag  (t_2 )\left| 0 \right\rangle
	\end{eqnarray}
We would integrate these diagrams over all space and over time from 0 to $\infty$.  But we will observe that $G_0$ is zero for negative times $t_1-t_2$, and so this will allow us to integrate over all time as well without error.  

\subsection{Simplification of the diagrammatic expansion}

The fact that the expectation is taken against the vacuum has a few important consequences for the diagrammatic expansion.  First we will find that we have no self-energy corrections to $G_0$.  This is because the self-energy is due to the interaction of the particle with the background state.  But since there are no particles in our background state, the self-energy vanishes.  Also, because $G_0$ does not allow negative times, all diagrams must proceed forward in time.  This will rule out all crossed diagrams, vertex corrections, etc.  The vacuum expectation dramatically simplifies the diagrammatic expansion, and we can determine its general form from the following argument.  Taking $N=3$ for illustrative purposes (the argument for general N is completely analogous), $G_3$ can be constructed via the expansion in Fig. \ref{Dyson-expansion}.  

	\begin{figure}
	\includegraphics[angle=0,width=0.47\textwidth]{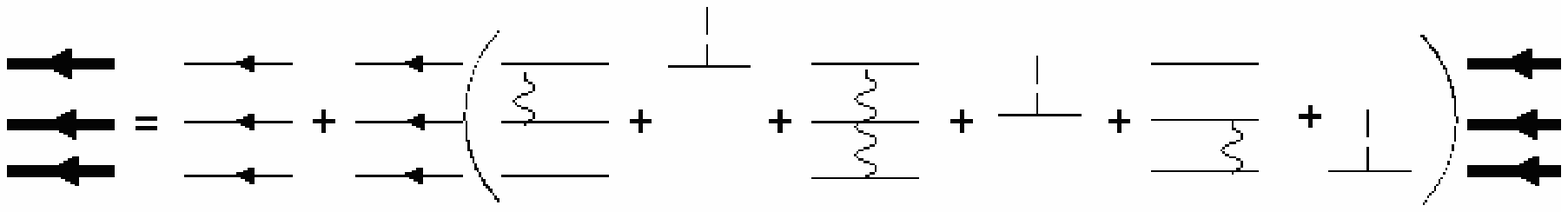}
	\caption{Dyson expansion of $G_3$. $G_3$ is represented by the bold lines.}
	\label{Dyson-expansion}
	\end{figure}

The proof is straightforward.  Simply write out the expansion algebraically, and then act on both sides of the equation with the operator
$-\partial/\partial t - H_0$, where $H_0$ is the kinetic energy term. The result will be the equation
	\begin{eqnarray}
	\left[ { - \frac{\partial }{{\partial t}} - H_0 - \sum\limits_{i = 1}^3 {u(x_i )}  - \sum\limits_{i < j}^3 {\upsilon (x_i     ,x_j )} } \right]G_3  = \cr 
	\delta (t)\delta (x_1 ,x_2 ,x_3 |y_1 ,y_2 ,y_3 )
	\label{Dyson-expasion-eq}
	\end{eqnarray}
which is indeed the desired differential equation for times greater than 0.

\subsection{Exponential expansion}

We will find it most convenient to analyze the series expansion of the Green's function in terms of an exponential series (defined below), rather than a Taylor series.  One reason is that we will find the individual diagrams in the Taylor expansion diverge logarithmically (for our potentials) for small y.  Yet when exponentiated, the diagrams will give logical, finite results.  Additionally, we will need to extract from the Taylor expansion the non-perturbative symmetry required in Eq.~(\ref{symmetry}), and this is perhaps most directly accomplished using the exponential series.  To that end it is useful to consider the following.  Each diagram in the Taylor expansion of $G_N$ can be labelled according to the number of single particle potential lines on each particle, and the number of interaction lines between each pair of particles. For convenience we will then label each diagram, $W$, in the Taylor expansion according to the following notation:
	\[
	\\W^{\{u_i\}}_{\{v_j\}}.
	\]
The superscripts, $\{u_i\}$, label the number of external potential lines on each particle, $i = 1...N$.  The subscripts $\{v_j\}$ label the number of interaction lines between each pair of particles $j = 1...N(N-1)/2$.  For instance, the diagram in Fig.\ref{Worder} 
	\begin{figure}
	\includegraphics[angle=0,width=0.30\textwidth]{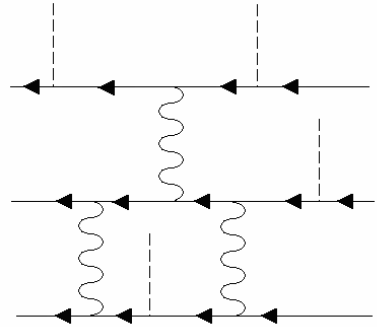}
	\caption{Typical diagram, W, in Taylor expansion of $G_N$}
	\label{Worder}
	\end{figure}
would be labelled $W^{211}_{102}$ since it has two u's on particle 1, one u on particle 2, and one u on particle 3, as well as one v between the first particle pair (12), 0 v's between the second particle pair (13), and two v's between the third particle pair (23).  Other diagrams obtained by rearranging the potential lines on this diagram would be of the same order.  Using this notation, the Taylor series expansion of the N-particle Green's function could be written as
	\begin{equation}
	G_N  = G_N^{(0)} + 	\sum\limits_{\{u_i\}  = 0}^\infty  \sum\limits_{\{v_j\}  = 0}^\infty 
  {W_{\{v_j\} }^{\{u_i\} } }
  \label{Tay-expansion}
  \end{equation} 
where $G^{(0)}_N$ is the unperturbed N-particle Green's function.  In preparation for the next step, we'll divide through by $G^{(0)}_N$ to get
	\begin{equation}
	\frac{{G_N }}{{G_N^{(0)} }} = 1 + \sum\limits_{\{u_i\}= 0}^\infty  \sum\limits_{\{v_j\}  =
	0}^\infty  {w_{\{v_j\}}^{\{u_i\}} }  
	\end{equation}
where we define the dimensionless diagram $w=W/G^{(0)}_N$.  Now we'd like to re-express this Taylor series as an exponential series.  So we write,
	\begin{equation}
	\frac{{G_N }}{{G_N^{(0)} }} = \exp \left[ {\sum\limits_{\{u_i\}= 0}^\infty  {\sum\limits_{\{v_j\} 0}^\infty  {f_{\{v_j\} }^{\{u_i\}} } } } \right].
	\end{equation}
To define $f$ with respect to $w$ we equate both expressions and solve for $f$ in terms of $w$ order by order, where 'order' is understood in the sense of the subscripts/superscripts $^{u_1,...,u_N}_{v_1,...,v_{N(N-1)/2}}$.  This process in general defines $f$ in the exponent.  Formal rules can be worked out and they are as follows:

The N-particle Green's function is given by:
	\begin{equation}
	G_N  = G_N^{(0)} \exp \left[ {\sum\limits_{\{u_i\}= 0}^\infty  {\sum\limits_{\{v_j\} = 	
	0}^\infty  {f_{\{v_j\} }^{\{u_i\} } } } } \right]
	\label{G_Nform}
	\end{equation}
where the superscripts $\{u_i\}$ label the number of external potential lines on each particle with $i = 1...N$, and  the subscripts $\{v_j\}$ label the number of interaction lines between each pair of particles with $j = 1...N(N-1)/2$.  There is a one-to-one correspondance between a particular dimensionless diagram, $w$, in the Taylor series expansion of $G_N$, and a particular term $f$ in the exponential series expansion.  The rules for constructing the $f$ corresponding to the $w$ are as follows:
\begin{itemize}
\item[1.] Start with the basic diagram $w^{\{u_i\}}_{\{v_j\}}$
\item[2.] Vertically cut $w^{\{u_i\}}_{\{v_j\}}$ inbetween interaction and potential lines, into all distinct diagram 
		 factorizations (daughter diagrams) such that the sum of the corresponding indices in each daughter
	   diagram adds up to: $^{\{u_i\}}_{\{v_j\}}$
\item[3.] For each daughter diagram, each $w$ receives a (-) sign.
\item[4.] For each daughter diagram, let $n$ be the number of $w$'s in the daughter diagram.
		 Then divide by $n$.  
\end{itemize}

To illustrate, consider the particular diagram in the Taylor series expansion in Fig.~(\ref{Wparent}) (for N = 3).  
	\begin{figure}[h]
	\includegraphics[angle=0,width=0.30\textwidth]{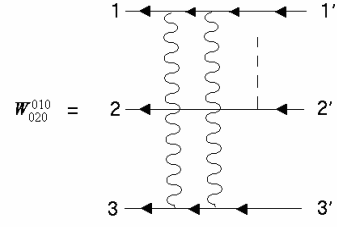}
	\caption{Typical diagram, W, in Taylor expansion of $G_3$}
	\label{Wparent}
	\end{figure}
We can cut this diagram in 3 different ways.  One way is between the two interaction lines in Fig.~(\ref{Wdaughter1}), 
	\begin{figure}[h]
	\includegraphics[angle=0,width=0.45\textwidth]{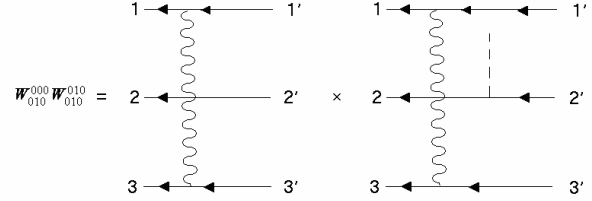}
	\caption{Daughter diagram 1}
	\label{Wdaughter1}
	\end{figure}
between the last interaction line and the external potential line in Fig.~(\ref{Wdaughter2}),
	\begin{figure}[h]
	\includegraphics[angle=0,width=0.45\textwidth]{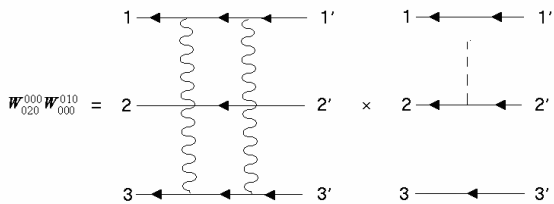}
	\caption{Daughter diagram 2}
	\label{Wdaughter2}
	\end{figure}
and between both, Fig.~(\ref{Wdaughter3}),

	\begin{figure}[h]
	\includegraphics[angle=0,width=0.45\textwidth]{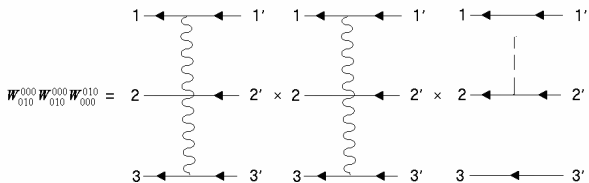}
	\caption{Daughter diagram 3}
	\label{Wdaughter3}
	\end{figure}
resulting in the following expression for $f$:  
	\bea
	f_{020}^{010}  &=&  - w_{020}^{010}  + \frac{1}{2}w_{010}^{000} w_{010}^{010}  + \frac{1}{2}w_{020}^{000}
	w_{000}^{010}  \cr
	&-& \frac{1}{3}w_{000}^{010} w_{010}^{000} w_{010}^{000} .
	\eea
We recall that $w = W/G_3$ in this case.  

\subsection{Extraction of symmetry-satsifying component of $G_N$}

There is a set of $f$'s appearing in the exponential series of $G_N$ that we can sum exactly.  These are the ones which give us the exact single particle Green's function, $G_1$, and two-particle Green's function $G_2$.  It turns out that these terms are also responsable for the asymptotic behavior in Eq.~(\ref{symmetry}).  Going back to Eq.~(\ref{G_Nform}), consider the set of diagrams parameterized by $f^{u_10...0}_{0...0}$.  This set has $u_1$ external potential lines on particle one, and nothing else on the other N-1 particle lines.  If we add these diagrams up, then we get:
	\begin{eqnarray}
	\exp \left\{ {\sum\limits_{\left\{ {v_j } \right\} = 0}^\infty  {\sum\limits_{\left\{ {u_j } \right\} = 0}^\infty 
	{f_{00...0}^{u_10...0} } } } \right\} &=& 1 + \sum\limits_{\left\{ {v_i } \right\} = 0}^\infty  {\sum\limits_{\left\{
	{u_j } \right\} = 0}^\infty  {w_{00...0}^{v_10...0} } }\cr
	&=& \frac{{G_1 (1)G_{N - 1}^0 }}{{G_N^0 }}\cr
	&=&\frac{{G_1 (1)}}{{G_1^0 (1)}}.
	\label{G1factor}
	\end{eqnarray}
Similarly, adding up $f^{0u_2...0}_{00...0}$ would give us $G_1(2)/G_1^{0}$.  Now consider the set of diagrams given by $f^{00...0}_{v_10...0}$.  This would correspond to the set of interaction ladders between the first pair of particles, say particles 1 and 2.  And there would be no interaction lines between any other pair of particles (and no external potential lines for that matter).  If we add these up, we will obtain
	\begin{eqnarray}
	\exp \left\{ {\sum\limits_{\left\{ {v_j } \right\} = 0}^\infty  {\sum\limits_{\left\{ {u_i } \right\} = 0}^\infty 
	{f_{v_10...0}^{00...0} } } } \right\} &=& 1 + \sum\limits_{\left\{ {v_j } \right\} = 0}^\infty  {\sum\limits_{\left\{
	{u_i } \right\} = 0}^\infty  {w_{v_10...0}^{00...0} } }\cr
	&=& \frac{{G_2 (1,2)G_{N - 2}^0 }}{{G_N^0 }}\cr
	&=&\frac{{G_2 (1,2)}}{{G_2^0 (1,2)}},
	\label{G2factor}
	\end{eqnarray}
and similarly for $f^{00...0}_{0v_2...0}$, etc.  Therefore let us separate the $f$-terms with only one non-zero index from the rest, which we'll denote by $f'$.  After adding up the one-index $f$'s we will obtain,

	\begin{eqnarray}
  G_N&=&G_N^{(0)} \exp \left[ {\sum\limits_{\{ v_j \}  = 0}^\infty  {\sum\limits_{\{ u_i \}  = 0}^\infty  {f_{\{ v_j
  \} }^{\{ u_i \} } } } } \right]\cr
  &=&G_N^{(0)} \prod\limits_i {\frac{{G_1 (i)}}{{G_1^0 (i)}}} \prod\limits_{i < j} {\frac{{G_2 (i,j)}}{{G_2^0 (i,j)}}}\cr
  &\times&
  \exp \left[ {\sum\limits_{\{ v_j \}  = 0}^\infty  {\sum\limits_{\{ u_j \}  = 0}^\infty  {f'\,_{\{ v_j \} }^{\{ u_j \}
  } } } } \right].
	\label{G_Nfinal}
	\end{eqnarray}

This is our main result for the Green's function, $G_N$.  Again, $f'$ consists of all diagrams which don't consist purely of external potential lines on one particle, or interactions lines between one pair of particles.  The first order contribution to $f'$ will be considered in the Appendix.    

On a general note, the unperturbed basis, $H_0$, does not have to be the free particle basis; any basis will suffice.  Given a choice for $H_0$, $H_0$ will define $G_N^{(0)}$, and will be the basis in which the diagrams in $f'$ are expanded.  $G_1$ would be defined via $H_0+\delta u$, where $\delta u$ is the single particle perturbation. 
$G_2$ would be defined via $H_0 + v$, where $v$ is the two-particle interaction.  For instance, if the single particle potential is simple enough, then the exact single particle basis may be convenient to use.  In that case, there would be no external potential lines appearing in the the $f'$ diagrams, only interaction lines.  

It will be evident that the terms to the left of the exponential in Eq.~(\ref{G_Nfinal}) possess the symmetry present in Eq.~(\ref{symmetry}), and as such would constitute a natural 'first order' approximation, though they contain contributions up to infinite order in both $\lambda_u$ and $\lambda_v$.  The $f'$ terms in the exponent are corrections to this approximation.  As evidence of the adequacy of this first order approximation, we adduce the fact that by itself, it reproduces the solution of the Q1D DMPK equation for the distinctly non-perturbative values of $\gamma = 1,2,4$.  This in itself is remarkable, since the DMPK equation is a many-body problem, but the solution, at least in the metallic and insulating regimes, can apparently be factored into 1 and 2 body Green's functions.  Thus, the first order approximation may be more generally accurate than it might seem.  Indeed, the first order approximation treats all one and two particle correlations exactly to infinite order, and approximates all higher order correlations as products of these lower order correlations.  The $f'$ diagrams correct this approximation.  So if three particle correlations are important then the $f'$ terms will need to be included.  But if not, then we may expect that our 'first' order approximation may be highly accurate - even for non-perturbative interactions.  

\section{Distribution of the eigenvalues}
Now we would like to apply this formalism to our problem of calculating $p(\textbf{x},t|\textbf{y})$.  To wit, we need to calculate the 1 and 2 particle Green's functions.  Since the single particle potential $u(x)$ is not small, we would like to include it in our basis Hamiltonian, $H_0$.  

\subsection{Single particle Green's function}
$G^{(0)}_1$ will satisfy the differential equation,
	\begin{equation}
	- \frac{\partial }{{\partial t}}G_1^{(0)}  = \left[ { - \frac{K}{4}\frac{{\partial ^2 }}{{\partial x^2 }} + u(x)}
	\right]G_1^{(0)}
	\end{equation}
This equation has been solved in the literature \cite{beenakker}.  The un-normalized eigenfunctions are $\psi_k(x) = \sinh x P_\nu^0(\cosh x)$ where $\nu = -1/2+ik$, and $P^0_\nu (x)$ is a Legendre function.  The small/large x behavior of the eigenfunctions is given by
	\begin{eqnarray}
  \label{G_1ass}
  \psi_k (x) &\sim&  x^{1/2},  \;\;\; x < < 1\cr
  \psi_k(x) &\sim& \sqrt {\tanh x} {\mathop{\rm Re}\nolimits} \left[ {\frac{{\Gamma
  (ik)}e^{ikx}}{{\Gamma (ik + \gamma /2)}}} \right], \;\;\; x > 1.
	\end{eqnarray}
For large times (lengths) $t=L_z/\ell$, only the lowest lying eigenstates $(k < < 1)$ are important to the Green's function, and we will make this approximation on the $\Gamma$ function so that $\Gamma (ik)/\Gamma (1/2+ik) \approx 1/ik$.  Normalizing the eigenfunction and evaluating the Green's function we obtain, in the large x/small y limit:
	\begin{equation}
	G_N^0 ({\bf{x}},t|{\bf{y}} <  < 1) = \prod\limits_{i = 1}^N {\frac{1}{{\sqrt {K\pi t} }}e^{ - x_i^2 /Kt} x_i
	y_i^{1/2} } .
	\end{equation}

\subsection{Two-particle Green's function}

Next we need the two-particle Green's function.  Since we're using the exact single particle basis, the external potential $u(x)$ will appear in the equation for $G_2$,
	\begin{equation}
  - \frac{\partial }{{\partial t}}G_2  = \left[ {\sum\limits_{i = 1}^2 {\left( { - \frac{K}{4}\frac{{\partial ^2
  }}{{\partial x_i }} + u(x_i )} \right)}  + \upsilon (x_1 ,x_2 )} \right]G_2 .
  \end{equation}
This equation would be rather difficult to solve, so we will use an approximation afforded to us in our region of interest. Note above in Eq.(\ref{G_1ass}), that $\psi_k(x) \sim \sin (kx)$ for $x > 1$ and $k<<1$.  Since we're concerned with the insulating state behavior, we will naturally be considering only large x.  Therefore we may replace the external potential $u(x)$ with the boundary condition that $\psi_k(x) = \sin (kx)$.  Given this, we can evaluate the $H_0$ two-particle Green's function in the large x, small y limit,  
	\begin{equation}
	G_2^{(0)} ({\bf{x}},t|{\bf{y}} <  < 1)\, \approx \frac{2x_1 x_2 y_1 y_2}{{\pi (Kt)^3 }}\exp \left[ { - \frac{{x_1^2  + x_2^2
	}}{{Kt}}} \right].
	\end{equation}
$G_2$ will then be defined through
	\begin{equation}
	- \frac{\partial }{{\partial t}}G_2  = \left[ { - \frac{K}{4}\frac{{\partial ^2 }}{{\partial x_1^2 }} -
	\frac{K}{4}\frac{{\partial ^2 }}{{\partial x_2^2 }} + \upsilon (x_1 ,x_2 )} \right]G_2
	\end{equation}
such that we use $\sin(kx)$ as the free basis. This equation can now be solved.  We first change variables to $z_1 = x_1 - x_2$ and $z_2 = x_1 + x_2$.  Then the equation is separable, and may write $G_2 = g(z_1)g(z_2)$ where $g(z)$ satisfes,
	\begin{equation}
	- \frac{\partial }{{\partial t}}g = \left[ { - \frac{K}{4}\frac{{\partial ^2 }}{{\partial z^2 }} +
	\frac{K}{4}\frac{{\gamma (\gamma  - 2)}}{2}\frac{1}{{\sinh ^2 z}}} \right]g.
	\end{equation}
Analogous to the single particle case, un-normalized eigenfunctions are given by
	\begin{eqnarray}
  \psi _k (z)\sim z^{\gamma /2},  \;\;\;\;\;\;\;\; z <  < 1 \cr
  \psi _k (z)\sim\sqrt {\tanh z} {\mathop{\rm Re}\nolimits} \left[ {\frac{{\Gamma (ik)}}{{\Gamma (ik + \gamma
  /2)}}e^{ikz} } \right], &  & z > 1.
 	\end{eqnarray}
To evaluate $g(z)$ in the insulating limit, we let $k < < 1$ and $\gamma < < 1$ as well.  With this approximation, $\Gamma(ik)/\Gamma(ik+\gamma/2)\approx(ik+\gamma/2)/ik$, and upon normalization of the $\psi$, we can construct the green's function $g(z)$.  Now the Green's function $g(z_1)g(z_2)$ does not satisfy the effective boundary conditions imposed by $u(x)$.  To satisfy these, we can add to this solution Green's functions corresponding to 'image charges' outside the boundary $x_1, x_2 = 0$.  Then we will have:
	\begin{eqnarray}
	&&G_2 ({\bf{x}},t|{\bf{y}}) \cr
	&=& g(x_1  - x_2 ,t|y_1  - y_2 )g(x_1  + x_2 ,t|y_1  + y_2 )\cr
	&-&g(x_1  + x_2 ,t|y_1  - y_2 )g(x_1  - x_2 ,t|y_1  + y_2 )\cr
	&-&g( - x_1  - x_2 ,t|y_1  - y_2 )g( - x_1  + x_2 ,t|y_1  + y_2 )\cr
	&+&g( - x_1  + x_2 ,t|y_1  - y_2 )g(x_1  - x_2 ,t|y_1  + y_2 ).
	\end{eqnarray}
Finally, forming the ratio $G_2/G_2^{(0)}$ and taking the small y limit, we obtain,
	\bea
	\label{G_2ass}
	\frac{{G_2 ({\bf{x}},t|{\bf{y}} <  < 1)}}{{G_2^0 ({\bf{x}},t|{\bf{y}} <  < 1)}} &\approx& T_\gamma  (x_1  + x_2
	)T_\gamma  (|x_1  - x_2 |)\cr
	&\times& |y_1^2  - y_2^2 |^{\gamma /2}
	\eea
where $T_\gamma(x)$ is defined as:
	\begin{equation}
	T_\gamma  (x) \equiv 1 - \frac{\gamma \sqrt {Kt} }{\sqrt{2}}{\mathop{\rm erfcp}\nolimits} \left( {\frac{{\gamma \sqrt {Kt}
	}}{2\sqrt{2}} + \frac{x}{\sqrt {2Kt} }} \right),
	\end{equation}
and we have defined
	\begin{equation}
	{\mathop{\rm erfcp}\nolimits} (x) \equiv \frac{{\sqrt \pi  }}{2}e^{x^2 } {\mathop{\rm erfc}\nolimits} (x).
	\end{equation}
Left out of Eq.~(\ref{G_2ass}) are the additional terms coming from the effect of the boundaries.  These also have the appropriate symmetry, but are negligible when $x > 1$.  Ultimately, we could have ignored the external potential, as we did in our previous paper,\cite{dm-09}, for purposes of calculating $G_2$, since it has effect only near the origin, but in the insulating regime, only $x >> 0$ is of consequence.  

\subsection{N-particle Green's function and $p(\textbf{x},t)$}

From $G_1$ and $G_2$ and using Eq.~(\ref{G_Nfinal}) we can construct the approximate N-particle Green's function to first order,  
	\begin{eqnarray}
	G_N  \approx \prod\limits_{i = 1}^N {\frac{1}{{\sqrt {K\pi \,t} }}e^{ - x_i^2 /Kt} x_i }\times \cr
	\prod\limits_{i < j}{T_\gamma  (x_i  + x_j )T_\gamma  (|x_i  - x_j |)y_i^{1/2} |y_i^2  - y_j^2 |^{\gamma /2} } .
	\end{eqnarray}
Observe that we have the requisite symmetry in the small y limit, as required by Eq.~(\ref{symmetry}).  Now from this we may construct $p(\textbf{x},t)$ using Eq.(\ref{singular}).  We obtain,
	\begin{eqnarray}
	p({\bf{x}},t) = \prod\limits_{i = 1}^N {\frac{1}{{\sqrt {K\pi \,t} }}e^{ - x_i^2 /Kt} x_i \sqrt {\sinh 2x_i }
	}\times \cr
	\prod\limits_{i < j} {|\sinh ^2 x_j  - \sinh ^2 x_i |^{\gamma /2} T_\gamma  (x_i  + x_j )T_\gamma  (|x_i  - x_j |)}.\cr
	\end{eqnarray}

\subsection{Deeply insulating limit of the smallest eigenvalue}

We can write the eigenvalue distribution in the form of a classical partition function. First recall that the 'time' variable is the longitudinal length of the conductor $t = L_z/\ell$.  So we have,
	\begin{equation}
	\frac{1}{Kt}=\frac{\xi}{4\ell}\frac{\ell}{L_z}=\Gamma .
	\end{equation}
Now we may write p(\textbf{x}) as,
	\begin{equation}
	p({\bf{x}}) \propto \exp \left[ { - \sum\limits_i {U(x_i )}  - \sum\limits_{i < j} {V(x_i ,x_j )} } \right]
	\end{equation}
where we have defined:
	\begin{eqnarray}
	U(x_i ) &=& \Gamma x_i^2  - \ln x_i  - \frac{1}{2}\ln \sinh 2x_i\cr
	V(x_i ,x_j ) &=&  - \frac{\gamma }{2}\ln |\sinh ^2 x_i  - \sinh ^2 x_j | \cr
	&-& \ln \left[ {T_\gamma  (x_i  + x_j)T_\gamma  (|x_i  - x_j |)} \right].
	\end{eqnarray}
In the insulating limit, $\gamma < < 1$, the $x_i$'s are large, and the conductance is dominated by the smallest eigenvalue.  Therefore the conductance is given by $g = \sum {1/\cosh^2x_i}\approx 1/\cosh^2x_1$.  In this case, we can obtain the full distribution $P(g)$ by considering the contribution from $x_1$ alone.  We can approximately obtain $P(x_1)$ via the following procedure.  We order the eigenvalues from least to greatest, separate out $x_1$, and then integrate over the rest,  
	\begin{equation}
	\label{px1}
	p(x_1 ) \propto e^{ - U(x_1 )} Z_{N - 1} 
	\end{equation}
where
	\begin{equation}
	Z_{N - 1}  = \int\limits_{x_1 }^\infty  {dx_2 \int\limits_{x_1 }^\infty  {dx_3 ...\int\limits_{x_1 }^\infty  {dx_N
  } } } p_{N - 1} ({\bf{x}})
  \end{equation}
and
	\bea
	p_{N - 1} ({\bf{x}}) &=& \exp [  - \sum\limits_{i = 2} {\left[ {U(x_i ) + V(x_1 ,x_i )} \right]}\cr
	&  -&
	\sum\limits_{2 \le i < j} {V(x_i ,x_j )}  ].
	\eea
We can consider $Z_{N-1}$ to be the partition function of a classical gas of $N-1$ particles constrained between $x_1$ and $\infty$.  Calculating this classical partition function is non-trivial in itself.  Though the interaction is weak, a cluster expansion will not work because the interaction is long-ranged; the individual terms in the expansion will diverge with the particle number.  Barring a more accurate method, we will use the mean-field approximation \cite{Chaiken}.  The basic idea is that 
	\begin{eqnarray}
	Z_{N - 1}  &=& e^{ - F_\rho  } \cr
	F_\rho   &=& \left\langle E \right\rangle _\rho   - \left\langle {\ln \rho ^{ - 1}
	} \right\rangle _\rho
	\end{eqnarray}
where $\langle E \rangle_\rho$ is the expectation of the energy of these particles with respect to $\rho$, $\langle \ln \rho^{-1} \rangle_\rho$ is the entropy, and  $\rho$ is the probability distribution function for the $N-1$ particles that minimizes the free energy $F_\rho$.   The mean-field approximation consists of postulating a trial $\rho$:
	\begin{equation}
	\rho (x_2 ,x_3 ,...,x_N ) = \prod\limits_{i = 2}^N {\frac{{n(x_i )}}{{N - 1}}}
	\end{equation}
where the $n(x)$ is normalized to $N-1$, and then minimizing $F_\rho$ with respect to $n$.  Forming $F_\rho$ with our trial $\rho$, we obtain,
	\begin{eqnarray}
	F_\rho   &=& \int\limits_{x_1 }^\infty  {dx\,n(x)} \left[ {U(x) + V(x_1 ,x)} \right]\cr
	&+& \frac{1}{2}\int\limits_{x_1}^\infty  {dx\int\limits_{x_1 }^\infty  {dy\,n(x)n(y)\,V(x,y)} } \, \cr
	&+& \int\limits_{x_1 }^\infty  {dx\,n(x)\ln\frac{{n(x)}}{{N - 1}}}.
	\end{eqnarray}
To find the $n(x)$ which minimizes $F$, we take a functional derivative with respect to $n$ and set it to 0, subject to the constraint that $n(x)$ be normalized to $N-1$.  We obtain the following equation,
	\begin{equation}
	\int\limits_{x_1 }^\infty  {dy\,V(x,y)n(y)}  + U(x) + V(x_1 ,x) + \ln \frac{{n(x)}}{{N - 1}} - \Lambda  = 0
	\end{equation}
where $\Lambda$ is the Lagrange multiplier enforcing the normalization constraint.  We now turn to evaluating the eigenvalue density, $n(x)$.  One can numerically solve this equation, but it is possible to make reasonable approximations based on the qualitative picture that for $x$ close to $x_1$, the external potential determines the form of the density while for large $x$, the interaction dominates, crystallizing the eigenvalues into a kind of lattice.  Given the fact that we expect $x_1$ to be large, we may make the following approximation on the interaction.  First, we may neglect the $T_\gamma$ interaction as it is rather small in comparison to the $\ln\sinh^2$ interaction.  Secondly, we may make the approximation that $\ln(\sinh^2x_i-\sinh^2x_j) \approx 2\max(x_i,x_j)$.  On the external potential, we may make the following saddle point approximation:
	\begin{equation}
	U(x) + V(x,x_1 ) \approx U_{sp} (x) = U'_{sp}  + \Gamma '(x - x'_{sp} )^2
	\end{equation}
where,
	\begin{eqnarray}
	U'_{sp}  &\approx&  - \frac{1}{{4\Gamma }} + \ln 2\Gamma \cr
	\Gamma ' &\approx& \Gamma  - 2\Gamma ^2 \cr
	x'_{sp}  &\approx& \frac{1}{{2\Gamma }} + 1.
	\end{eqnarray}
All of these approximations are quite good in the insulating regime.  Following this, our density equation becomes,
	\bea
	&&\gamma \int\limits_{x_1 }^x {dy\,x\,n(y)}  + \gamma \int\limits_x^\infty  {dy\,y\,n(y)}\cr
	&&+ U_{sp} (x) + \ln\frac{{n(x)}}{{N - 1}} + \Lambda  = 0
	\eea
This non-linear integral equation can now be reduced to a first order non-linear differential equation.  To do so, we will first need the following boundary conditions imposed by the integral equation itself.  For $x$ close to $x_1$, the entropy ($\ln$) term will dominate over the interaction (integral) term and we'll have,
	\begin{equation}
	\label{boundary1}
	 n(x)\sim\exp \left[ { - U_{sp} (x)} \right], \;\;\;\;\;\;\; x \to x_1. 
	\end{equation}
For large $x$, the interaction will dominate over the entropy, and we'll have:
	\begin{equation}
	\label{boundary2}
	n(x)\sim\frac{{2\Gamma '}}{\gamma }\;\;\;\;\;\;\; x \to \infty 
	\end{equation}
Now we take two derivatives of our integral equation for $n(x)$.  This results in the 2nd order non-linear differential equation:
	\begin{equation}
	- n''(x) + \frac{{n'(x)^2 }}{{n(x)}} - 2\Gamma 'n(x) + \gamma n(x)^2  = 0 .
	\end{equation}
Happily, this equation is autonomous in $x$, so we can change the independent variable to $n$, and the dependent variable to $p = dn/dx$.  This will give a Bernoulli differential equation in $p$.  Using the boundary condition in Eq.~(\ref{boundary2}), we can obtain the following differential equation for $n$.
	\begin{equation}
	\frac{{dn}}{{dx}} = n\sqrt {2\gamma n - 4\Gamma '\ln \frac{n}{{2\Gamma '/e\gamma }}} .
	\label{n(x)ODE}
	\end{equation}
The full solution to this equation is unknown, but in the insulating state, it is the small $|x-x_1|$ behavior that is important since only the lowest lying eigenvalues control the conductance.  For $x$ close to $x_1$, and therefore small $n$, the $\ln(n)$ term dominates the $n$ term in the square root.  Neglecting the linear term, we can perform the integral.  Employing the boundary condition Eq.~(\ref{boundary1}), we obtain,
	\begin{equation}
	n(x) \approx \frac{{2\Gamma '}}{{e\gamma }}e^{ - \Gamma '(x - x'_{sp} )^2 }, \;\;\; x \to x_1.
	\label{eq-density}
	\end{equation}

The numerical solution to this equation is shown in our previous paper \cite{dm-09} for two different values of $\Gamma$.  It illustrates that the density is roughly constant with an exponential gap from the origin that increases with increasing disorder.  In contrast, the density in the metallic regime is also constant, but starting at the origin.  Thus our result suggests that the opening of a gap in the eigenvalue spectrum could be considered as a signature of the metal-insulator Anderson transition.

Now that we have determined the density of eigenvalues for $x$ close to $x_1$, we can use it to evaluate $F_\rho$ in this same region.  We obtain,
	\begin{equation}
	F_\rho  (x_1 ) =  - \frac{{\sqrt \pi  }}{{8e\gamma \sqrt {\Gamma '} }}{\mathop{\rm erfc}\nolimits} \left[ {(x_1  -
	x'_{sp} )\sqrt {\Gamma '} } \right].
	\end{equation}
Finally, this brings us to our result for $p(x_1)$.  Using Eq.~(\ref{px1}), we obtain,
	\begin{eqnarray}
	p(x_1 ) &\propto& \exp \left[ { - f(x_1 )} \right]\cr
	f(x) &=& U(x) - \frac{{\sqrt \pi  }}{{8e\gamma \sqrt {\Gamma '} }}{\mathop{\rm erfc}\nolimits} \left[ {(x - x'_{sp}
	)\sqrt {\Gamma '} } \right].\cr
	\end{eqnarray}

\section{Distribution of conductance}

To obtain the probability distribution of the conductance in the insulating state, we use the aformentioned fact that the dominant contribution to $g$ comes from the first eigenvalue, $x_1$.  Therefore $\ln(g)\approx \ln(1/\cosh^2(x_1))$, which implies $x_1\approx (1/2)\ln(4/g)$.  And so,
	\begin{equation}
	\label{final}
	p(\ln g) \propto \exp \left[ { - f\left( {\frac{1}{2}\ln \frac{4}{g}} \right)} \right].
	\end{equation}
This is our expression for the conductance distribution in the insulating state.  In Fig.~\ref{P(g)graph}, we compare the analytical formula in Eq.~(\ref{final}) to numerical results and also to the distribution obtained from the GDMPK equation neglecting the interaction. 

	\begin{figure}[h]
	\includegraphics[angle=0,width=0.45\textwidth]{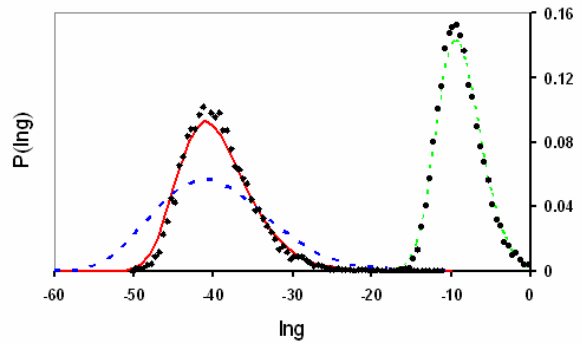}
	\caption{(Color online) $P(\ln g)$ in the insulating regime for two values of disorder, $\Gamma = 0.014$ (solid red
	line) and $\Gamma = 0.045$ (dotted green line), corresponding to $\langle \ln g\rangle = -39.4$ and $\langle \ln g
	\rangle = -8.9$, respectively.  The numerical data points are from Ref.~[\onlinecite{markos-unpublished} ] for the
	same value of $\langle \ln g \rangle$.  The dashed blue curve is the GDMPK prediction of $P(\ln g)$ for $\langle \ln
	g\rangle = -39.4$ neglecting the eigenvalue interaction}
	\label{P(g)graph}
	\end{figure}

Examining Eq.~(\ref{final}) we find the distribution has a Gaussian tail in $\ln g$ for very small $g$. We can calculate the expectation of $\ln g$ as a function of $\Gamma$.  The result, to leading order, is
	\begin{equation}
	\left\langle {\ln \left( {g} \right)} \right\rangle  \approx -\frac{1}{{\Gamma }}\left[ {1 -
	\sqrt {4\Gamma \ln \left( {\frac{1}{{4e\Gamma ^{3/2} }}} \right)} } \right].
	\label{lng_Gamma}
	\end{equation}
One may also calculate the variance of $\ln g$ in terms of $\langle\ln g\rangle$.  We find that asymptotically,
	\begin{equation}
	\\var(\ln g)\sim \frac{\langle \ln(1/g) \rangle}{\ln \langle \ln(1/g) \rangle}.
	\label{var_Gamma}  
	\end{equation}  
Current numerical simulations \cite{somoza}  suggest rather that the variance of $\ln g$ goes as $\langle -\ln g \rangle ^{1/2}$.  In our previous paper \cite{dm-09} we found a power law consistent with $2/3$.  It is a consequence of Eq.(\ref{var_Gamma}) that different power laws can be fitted over different ranges of $\ln g$.  To illustrate we plot $ var \ln g$ vs. $\langle -\ln g \rangle^{3/4}$ in Fig. (\ref{varlng}), and in the insert, over a shorter range of $\ln g$, we plot $ var \ln g$ vs. $\langle -\ln g \rangle^{2/3}$ which is actually a better fit over that shorter range. 
	\begin{figure}[h]
	\includegraphics[angle=0,width=0.45\textwidth]{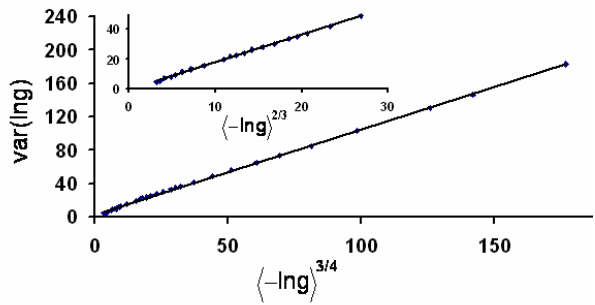}
	\caption{Variance of $\ln g $ plotted as a function of $\langle -\ln g \rangle^{3/4}$.  The points are calculated from
	Eq.(\ref{final}).  The line is a best fit.  Variance of $\ln g$ is plotted as a function of $\langle -\ln g  
	\rangle^{2/3}$ in the insert}
	\label{varlng}
	\end{figure}
So the discrepancy could be the result of the range of current numerical simulations (as the analytic solution seems to match with different power laws in different regimes), or the result of the series of approximations made in calculating the partition function, which provides motivation for further work in this area. 

A major prediction of Eq.~(\ref{final}) is the resulting novel disorder dependence of the skewness of the distribution. As shown in Ref.~[\onlinecite{dm-09}], the skewness is positive in the deeply insulating regime and seems to tend to a constant value as disorder increases.  From Eq.~(\ref{final}), one can calculate the skewness and confirm that it does approach a constant value, $\chi(\ln g) \sim  1.1$.  The convergence to the asymptote is logarithmic, as illustrated in Fig.\ref{skewness}.

	\begin{figure}[h]
	\includegraphics[angle=0,width=0.45\textwidth]{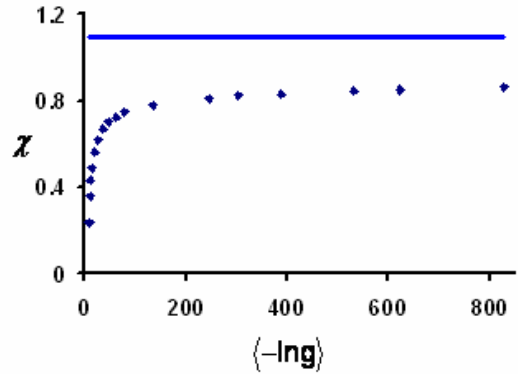}
	\caption{The skewness is plotted vs. $\langle -\ln g\rangle$.  Note the asymptote at $\chi \sim 1.1$.}
	\label{skewness}
	\end{figure}

As disorder decreases the skewness eventually changes sign while still in the insulating regime. This change of sign occurs well before the Anderson metal-insulator transition point. Unfortunately due to the approximations used to obtain the density of eigenvalues, we can not trust Eq.~(\ref{final}) to describe the distribution at the transition point itself.

\section{Summary and conclusion}

In this work we start with the assumption, well supported numerically, that the generalized DMPK equation (GDMPK) with two scale-dependent parameters correctly describes the evolution of transmission eigenvalues with increasing length in a 3D disordered conductor. Taking advantage of a mapping of the GDMPK equation on to an imaginary time Schr\"odinger equation for a system of bosons interacting via a short range interaction in 1D, we then develop a systematic perturbation theory to solve for the full distribution of the eigenvalues in terms of the $N$-particle Greens function describing the system (Eq.~(\ref{G_Nfinal})). In the strong disorder regime the strength of the  interaction between bosons is small, which we use as the small parameter in our perturbation theory. In particular, we make sure that the $N$-particle Greens function satisfies a highly non-trivial boundary condition. To leading order which includes at least all two-particle correlations, we obtain the distribution of the eigenvalues in the strong disorder regime, where all eigenvalues are exponentially large. Writing the eigenvalue distribution in the form of a classical partition function, we then separate out the smallest eigenvalue $x_1$ which dominates the distribution, and use a mean field approximation that allows us to evaluate the density of the eigenvalues (Eq. (\ref{eq-density})) and the free energy close to $x_1$. Finally, using Landauer formula connecting conductance with the transmission eigenvalues, we arrive at our expression for the full distriburtion of conductance, given by Eq. (\ref{final}). It turns out that the final distribution can be expressed in terms of a single disorder parameter.

We use Eq.~(\ref{final}) to compare directly with conductance distributions evaluted numerically from three dimensional tight binding Anderson model in the insulating regime. Once the parameter of the theory is fixed to give the correct mean value of the distribution, the entire distribution is reproduced very well. In addition, we evaluate the variance and skewness as a function of disorder. While the numerical data for the variance is usually fitted with power laws, we find that the asymptotic behavior is perhaps more complex, given by Eq.~(\ref{var_Gamma}), which can be fitted with different power laws in different regimes. We predict the skewness approaches a constant value in the deeply insulating regime, and changes sign as the disorder is decreased from deeply insulating regime towards the Anderson transition point. Our expression becomes less applicable as one approaches the transition, but the change in sign occurs well before the transition. 

The theory can in principle be improved by keeping higher order terms in the perturbation expansion, as shown in Appendix A. However, even though the perturbation parameter remains small up to the Anderson transition point, the partition function and hence the density of the eigenvalues can not be evaluated very accurately as the system approaches the transition. It is not clear at this point how to improve the calculation of the partition function beyond the mean field approximation used here. Nevertheless, starting at the strongly insulating regime, the present approach already predicts highly non-trivial changes in the conductance distribution as one decreases the disorder towards the Anderson transition.

\begin{appendix}
\section{Corrections to $G_N$}

If we want to include three-particle correlations, then to leading order they would come from the following type of diagram (Fig.~(\ref{threepart})) with two interaction lines.
	\begin{figure}[h]
	\includegraphics[angle=0,width=0.35\textwidth]{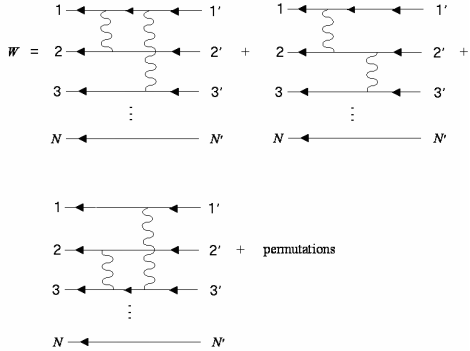}
	\caption{three particle correlation diagram}
	\label{threepart}
	\end{figure}
'Permutation' refers to the same graphs but with the order of the first and second interactions switched.  There would be the same set of diagrams for every distinct triple of particles within the set of $N$.  The first dimensionless graph would correspond to an $f$ of:
	\begin{equation}
	f_{110...0}^{00...0}  =  - w_{110...0}^{00...0}  + \frac{1}{2}w_{10...0}^{00...0} w_{01...0}^{00...0}.
	\end{equation}
Let us call this term $f_3(1,2,3;1',2',3')$.  Each of the other five graphs is simply a permutation of the arguments of the first.  For instance, switching the order of the interaction lines of $f_3$ would give us: $f_3(1',2',3';1,2,3)$.  The second $f$ would be $f_3(2,1,3;2',1',3')$, and its permutation would be $f_3(2',1',3';2,1,3)$.  The third $f$ would be $f_3(3,2,1;3',2',1')$ and its permutation would be $f_3(3',2',1';3,2,1)$.  Let $F_3(1,2,3)$ denote the sum of these 6 $f$'s.  Then using Eq.(\ref{G_Nfinal}) we can write out the expression for $G_N$ correct to second order in the interaction:
	\begin{equation}
	G_N \, \approx G_N^{(0)} \prod\limits_i {\frac{{G_1 (i)}}{{G_1^0 (i)}}} \prod\limits_{i < j} {\frac{{G_2
	(i,j)}}{{G_2^0 (i,j)}}\prod\limits_{i < j < k} {e^{F_3 (i,j,k)} } } .
	\end{equation}

In order to calculate these diagrams it is convenient to go to a plane-wave basis.  Unfortunately, the potentials in Eq. (\ref{potentials}) do not possess well-defined Fourier transforms due to the $1/x^2$ singularity at the origin. Even then, we have found that a good representation of the Fourier transform is given by the following relations:
	\begin{eqnarray}
	 u(q)=\lambda_u\pi q \coth(\pi q/4)\cr
	 v(q)=\lambda_v\pi q\coth(\pi q/2).
	 \end{eqnarray}
The large $q$ behavior of these potentials result in ln-divergences of the first order (and probably higher order too) dimensionless diagrams, $w$.  But the corresponding $f$'s result in well-defined, indeed necessary, terms by taking the divergences to the exponent.  This is one of the advantages of the exponential expansion, the divergences are resummed to give finite results order by order which we see accumulated in $G_1$ and $G_2$.  On the other hand, if one is not interested in the small $x_i$ or small $x_i-x_j$ behavior (this ought to be taken care of via $G_1$ and $G_2$ in any event), we may approximate these Fourier transforms by their small $q$ limit
	 \begin{eqnarray}
	 u(q)\approx 4\lambda_u\cr
	 v(q)\approx 2\lambda_v .
	 \end{eqnarray}
This is equivalent to approximating the potentials with a delta function which gives the same phase shift.  Since the potentials are very short ranged, this should be adequate.  Whichever set of expressions we use, the Feynman rules in frequency space would be given in Fig.~(\ref{G0}), and Fig.~(\ref{vinteraction}). 
	\begin{figure}
	\includegraphics[angle=0,width=0.35\textwidth]{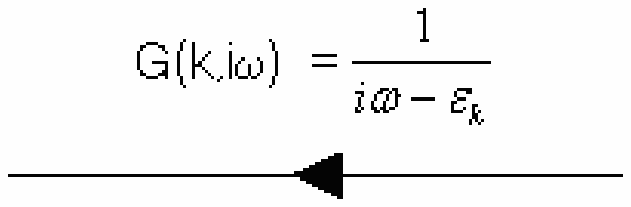}
	\caption{Unperturbed Green's function $G_0$ in Fourier space}
	\label{G0}
	\end{figure}
	
	\begin{figure}
	\includegraphics[angle=0,width=0.35\textwidth]{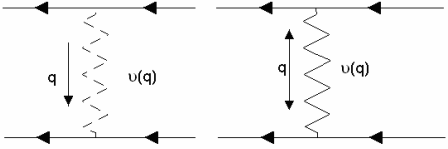}
	\caption{Interaction vertices in Fourier space}
	\label{vinteraction}
	\end{figure}
The first vertex corresponds to the interaction potential $v(q)=\lambda_v/\sinh^2(x_i-x_j)$, and the solid vertex corresponds to the interaction potential $v(q)=\lambda_v/\sinh^2(x_i+x_j)$.  The latter can be incorporated into Fourier space provided we associate it with momentum $q$ flowing from the center outwards towards both $x_1$ and $x_2$, rather than with momentum $q$ flowing from $x_1$ to $x_2$ as with the difference potential.  Momentum would be conserved at all intersections as usual.
\end{appendix}


\begin{thebibliography}{99}
\bibitem{note1} Note that this wave function is the exact wavefunction for a repulsive delta function potential of strength $\gamma$.
\end{thebibliography}
\end{document}